\documentclass[pre,superscriptaddress,twocolumn,showpacs,amsmath,amssymb,floatfix]{revtex4-1}
\usepackage{graphicx,bbm}

\def\etal#1{#1}
\def\tit#1{#1, }

\def\one{\mathbbm{1}}
\def\tr{\,{\rm tr}\,}

\def\ave#1{\langle #1 \rangle}
\def\ii{{\rm i}}
\def\sx{\sigma^{\rm x}}
\def\sy{\sigma^{\rm y}}
\def\sz{\sigma^{\rm z}}
\def\rrho{\tilde{\rho}}

\usepackage{hyperref}

\begin{document}
\title{Transport properties of a boundary-driven one-dimensional gas of spinless fermions} 

\author{Marko \v Znidari\v c}
\affiliation{Instituto de Ciencias F\' isicas, Universidad Nacional Aut\' onoma de M\' exico, Cuernavaca, M\' exico}
\affiliation{Physics Department, Faculty of Mathematics and Physics, University of Ljubljana, Ljubljana, Slovenia}
\author{Bojan \v Zunkovi\v c}
\affiliation{Physics Department, Faculty of Mathematics and Physics, University of Ljubljana, Ljubljana, Slovenia}
\author{Toma\v {z} Prosen}
\affiliation{Physics Department, Faculty of Mathematics and Physics, University of Ljubljana, Ljubljana, Slovenia}
\date{\today}

\begin{abstract}
We analytically study a system of spinless fermions driven at the boundary with an oscillating chemical potential. Various transport regimes can be observed: at zero driving frequency the particle current through the system is independent of the system's length; at the phase-transition frequency, being equal to the bandwidth, the current decays as $\sim n^{-\alpha}$ with the chain length $n$, $\alpha$ being either $2$ or $3$; below the transition the scaling of the current is $\sim n^{-1/2}$, indicating anomalous transport, while it is exponentially small $\sim \exp{(-\frac{n}{2\xi})}$ above the transition. Therefore, by a simple change of frequency of the a.c. driving one can vary transport from ballistic, anomalous, to insulating.  
\end{abstract}

\maketitle
\section{Introduction} 

A system of free fermionic particles is of great importance in many areas of physics. It can describe a gas of electrons, a solid state system in a tight-binding approximation, or the so-called XX spin chain, to name just a few. In a noninteracting case without fields physics is rather simple. Free modes, either in real or in momentum space, propagate without mutual interaction or dissipation with constant speed, resulting in ballistic transport properties. Things become more interesting in the presence of external fields. Limiting the discussion to one-dimensional (1D) systems, any on-site disorder results in a localization of all states -- the Anderson localization~\cite{Anderson}. A constant static electric field causes Bloch oscillations~\cite{Bloch} in a periodic lattice. An oscillating electric field on the other hand results in a renormalized hopping strength, leading to localization at specific values of the field or the frequency~\cite{fieldE}. In the present work we shall study a new setting, namely that of a harmonically (a.c.) boundary-driven system of free fermions, and show that, depending on the driving frequency, very different transport regimes can be observed. The model can also be seen as a minimal model for the boundary driven non-equilibrium quantum phase transition. We also discuss its experimental implementations, showing that it could be realized in an optical setting with cold atoms or ions or, in its fermionic version, with mesoscopic systems.

We shall study a 1D chain of spin-1/2 particles interacting between nearest-neighbor sites with an exchange interaction of the XX type. The Hamiltonian is
\begin{equation}
H=\sum_{j=1}^{n-1} (\sigma_j^{\rm x} \sigma_{j+1}^{\rm x} +\sigma_j^{\rm y} \sigma_{j+1}^{\rm y}).
\label{eq:H}
\end{equation}
Equivalently, using Jordan-Wigner transformation the XX model can be mapped to a system of non-interacting spinless fermions with the Hamiltonian $H=2\sum_k (c_k^\dagger c_{k+1} + c_k c_{k+1}^\dagger)$, where $c_k = \sigma^-_k\prod_{j<k}\sz_{j} $, and $c_k^\dagger= 
\sigma^+_k\prod_{j<k}\sz_{j}$ ($\sigma^\pm_k \equiv (\sx_k\pm \ii \sy_k)/2$) satisfy the standard fermionic algebra. The first and the last spin are coupled to magnetization reservoirs. In the language of fermions the reservoirs impose an external chemical potential determining the number of fermions in the system. The evolution of the system's density matrix is described by the Lindblad master equation~\cite{Lindblad},
\begin{equation}
{\rm d}\rho/{\rm d}t=\ii [ \rho,H ]+ {\cal L}^{\rm dis}(\rho),
\label{eq:Lin}
\end{equation}
where a dissipator ${\cal L}^{\rm dis}(\rho)=\sum_k \left( [ L_k \rho,L_k^\dagger ]+[ L_k,\rho L_k^{\dagger} ] \right)$ takes into account the influence of two reservoirs. The left reservoir is described by a pair of time-dependent Lindblad operators $L_{1,2}(t)=\sqrt{\varepsilon(1\pm \mu(t))}\,\sigma^\pm_1$, and similarly at the right end, $L_{3,4}(t)=\sqrt{\varepsilon(1\mp \mu(t))}\,\sigma^\pm_n$. Note that drivings at the two ends are opposite. For instance, when there is a maximal coefficient of $\sigma^+_1$ at the left end, we have a minimal coefficient of $\sigma^+_n$ at the right end. Two important bath parameters are the coupling strength $\varepsilon$ between the chain and the reservoirs and the value of the ``chemical potential'' that is oscillating in time, $\mu=\mu_0 \cos \omega t $. Nonzero $\mu$ means that probabilities for an injection and absorption of a spin/fermion are different. For d.c. driving, $\omega=0$, the transport of magnetization is trivially ballistic, i.e., the value of the current through the system is independent of the length $n$, and the nonequilibrium stationary solution of the Lindblad equation is known~\cite{Karevski:09} and can be compactly written in a matrix product operator form~\cite{JPA:10}. We shall show by a fully analytic calculation that the situation is rather different and interesting for $\omega \neq 0$.

\section{The Solution}

We want to find an asymptotic solution $\rho(t \to \infty)$ of the Lindblad equation to which the system converges after very long time. Because the system is harmonically driven this solution is not time-independent, as is the case for $\omega=0$, but instead oscillates with the very same forcing frequency $\omega$. In fact, the asymptotic stationary solution of Eq.~(\ref{eq:Lin}) can be written in the form 
\begin{equation}
\rho(t)=2^{-n}\one + \frac{1}{2}(\tilde{\rho}e^{\ii \omega t} + \tilde{\rho}^\dagger e^{-\ii\omega t}),
\end{equation} 
where $\rrho$ is time-independent. The nonequilibrium stationary state $\tilde{\rho}$ is in our case unique. There are several ways of computing $\rrho$. One is to realize that the XX model, even when time-dependent, belongs to a class of systems~\cite{temme:09,XXdephasing,Eisler:11} in which exponentially many equations for all $r$-point functions decouple into a hierarchy in which one has separate equations for each order of correlations. This greatly simplifies the Lindblad equation, enabling one to find the exact non-equilibrium steady state~\cite{XXdephasing} and even study time-evolution towards a steady state~\cite{Eisler:11}. We shall be especially interested in $2$-point correlations as these include local magnetization $\sz_k$ ($=$fermion density) as well as spin current ($=$particle current) $j_k=2(\sx_k \sy_{k+1} -\sy_k \sx_{k+1})$. In fact, the only nonzero $2$-point terms in $\rrho$ are $B^{(r)}_k=2(c^\dagger_k c_{k+r}-c_k c^\dagger_{k+r})$ for even $r$ and $B^{(r)}_k=2\ii(c^\dagger_k c_{k+r}+c_k c^\dagger_{k+r})$ for odd $r$ ~\cite{footnote1}. For instance, the first two operators in the series are $B^{(0)}_k=-2\sz_k$ and $B^{(1)}_k=j_k/2$. In a chain of length $n$ there are in total $n(n-1)/2$ such operators and one can write a self-contained set of as many equations for their unknown coefficients in $\rrho$. Due to our parametrization of driving they are all exactly proportional to $\mu_0$; from now on we therefore set $\mu_0=1$ in all our results. Using standard procedures such a set of linear equations can be solved for chain lengths of $n\sim 10^3$. Once these $2$-point coefficients are known, they can be used as inhomogeneous source terms in the set of equations for all $4$-point terms (which are all proportional to $\mu_0^2$), and so on (see~\cite{XXdephasing} for an example of such a calculation). 

In our case though one can do even better. 
Following \cite{prosen:njp,bojan:10,enej:11} the evolution of $2$-point correlations (i.e., the above mentioned $n(n-1)/2$ linear equations) can be compactly written in terms of a linear matrix equation. Defining the time dependent covariances $\langle w_jw_k\rangle\equiv \tr \rho(t)w_j w_k  = \delta_{j,k}-\ii Z_{j,k}$, where $w_{m}=c_m+c_m^\dag$, $w_{m+n}=\ii(c_m-c_m^\dag)$, $m=1,\ldots n$, and applying the Lindblad master equation, we obtain a differential equation for the $2n\times 2n$ covariance matrix ${\bf Z}$
\begin{equation}
\label{eq:diffZ}
{\rm d}{\bf Z}/{\rm d}t= -{\bf X}^{\rm T}{\bf Z}-{\bf ZX} + {\bf Y}\cos\omega t,
\end{equation}
where ${\bf X}=2\ii\,\sy\otimes{\bf J}-2\varepsilon\mathbbm{1}_2 \otimes{\bf R}$ and ${\bf Y}=-4\ii\,\sy\otimes{\rm \bf P}$ are real matrices, with ${\bf J}, {\bf P}, {\bf R}$ being $n\times n$ matrices having nonzero elements $J_{k,k+1}=J_{k+1,k}=1$, $k=1,\ldots,n-1$, and $R_{1,1}=R_{n,n}=P_{1,1}=-P_{n,n}=1$. By considering the symmetry of the differential equation (\ref{eq:diffZ}) we write the covariance matrix in the simple form ${\bf Z}(t)={\rm Re}\left(e^{\ii \omega t}(\mathbbm{1}_2\otimes{\bf Z}_0-\ii\,\sy\otimes{\bf Z}_2)\right)$. Inserting the last definition into (\ref{eq:diffZ}) we get two coupled equations for the $n\times n$ matrices ${\bf Z}_0$ and ${\bf Z}_2$. We immediately see that $({\bf Z}_0)_{j,k}=0$ for $j+k$ even, and $({\bf Z}_2)_{j,k}=0$ for $j+k$ odd, hence we can replace the two equations by one. Defining the orthogonal transformation $O_{j,k}=(-1)^{k+1}\delta_{j,k}$, the matrices ${\bf C}^{\pm}={\bf O}({\bf Z}_0\pm\ii {\bf Z}_2)$, and inverting the last relation, we end up with two uncoupled equations
\begin{equation}
\label{eq:lyap1}
2\{{\bf J} \pm {\rm i}\varepsilon {\bf R},{\bf C}^{\mp}\}\mp\omega{\bf C}^{\mp}=-4\varepsilon {\bf O P},
\end{equation}
where $\{,\}$ is an anticommutator. Note that  for odd $n$, ${\bf OP}={\bf P}$, while for even $n$, ${\bf OP}={\bf R}$. The last equation (\ref{eq:lyap1}) can be solved perturbatively in the coupling $\varepsilon$. The solution to first order in the coupling, obtained by the Fourier method (see the Appendix~\ref{sec:pert}), is
\begin{equation}
C_{j,k}^-=-32\varepsilon\!\!\!\!\!\!\!\!\sum_{\substack{p,m=1\\p+m=n ({\rm mod\,} 2)}}^n\!\!\frac{\sin{a_p}\sin{a_m}\sin{a_{jp}}\sin{a_{k m}}}{(n+1)^2(\lambda_p+\lambda_m)},
\label{eq:Cjk}
\end{equation}
with $\lambda_m=(\frac{\omega}{2}-4\cos{a_m})-\frac{8\ii\varepsilon}{n+1}\sin^2{a_m}$ where $a_k\equiv\frac{\pi k}{n+1}$.  It is instructive to emphasize the connection $C_{j,k}^-=-\ii\,(-1)^{j}\langle w_jw_k\rangle=-\frac{1}{2}(-1)^j \ave{B_j^{(|k-j|)}}$ if $j+k$ is odd and  $C_{j,k}^-=(-1)^{j}\langle w_{j+n}w_k\rangle=-\frac{\ii}{2}(-1)^j\ave{B_k^{(|k-j|)}}$ if $j+k$ is even. All expectation values in this work are computed with respect to the a.c. part of the density matrix $\rrho$, $\ave{A}=\tr{(\rrho A)}$. Note that $C_{j,k}^+=(-1)^{j+k+1}C_{j,k}^-$ and that $n\times n$ matrix ${\bf C}^-$ contains all non-vanishing matrix elements of $\bf{Z}$.

From the perturbative solution (\ref{eq:Cjk}) we can see~\cite{footnote3} that for large $n$ and $\omega>\omega_{\rm c}\equiv8$ the imaginary part of the Lyapunov equation (\ref{eq:lyap1}) can be neglected. In the thermodynamic limit one can in fact express the solution in terms of a generalized hypergeometric function (Appendix~\ref{app:hyper}). Here we prefer a different route by first taking a continuum limit and writing a partial differential equation. This results in a Helmholtz equation for the correlation function $C(x\equiv\frac{j}{n+1},y\equiv\frac{k}{n+1})=C_{j,k}^-$,
\begin{equation}
\nabla^2C(x,y)+(\omega_{\rm c}-\omega)n^2C(x,y)=P(x,y),
\label{eq:PDE}
\end{equation}
where $P(x,y)=\varepsilon(\delta(x-\frac{1}{n},y-\frac{1}{n})+(-1)^n\delta(x-1+\frac{1}{n},y-1+\frac{1}{n}))$ is the source term resulting from the driving. Since the dissipation effects in the matrix ${\bf X}$ can be neglected, the boundary conditions are $C(0,y)=C(x,0)=C(1,y)=C(x,1)=0$. Using the method of mirror images the problem can be unfolded into an infinite plane, where the source term becomes a grid of quadrupoles $P(x,y)\rightarrow \varepsilon \sum_{j,k=-\infty}^\infty \sum_{\nu=0}^1 (-1)^{\nu n}q(x-2j-\nu,y-2k-\nu)$. The solution is
\begin{equation}
C(x,y)=\varepsilon\!\!\!\!\!\sum_{j,k=-\infty}^{\infty}\sum_{\nu=0}^1 \frac{(-1)^{\nu n}}{n^2}\, G(x-2j-\nu,y-2k-\nu),
\label{eq:solC}
\end{equation}
where $G(x,y)$ is the Green's function of Eq.~(\ref{eq:PDE}) for a single quadrupole source $q(x,y) \equiv 2\delta'(x) \delta'(y)$ at the origin, where $\delta'(x)$ denotes the derivative of Dirac's function in the sense of distribution. For the critical frequency $\omega=\omega_{\rm c}$ the Green's function is particularly simple (and $n$-independent)
\begin{equation}
G_{\rm c}(x,y)=4xy/[\pi(x^2+y^2)^2].
\label{eq:prehod}
\end{equation}
For $\omega<\omega_{\rm c}$,  we can use heuristic arguments, combining integral representation of the sum (\ref{eq:Cjk}) and stationary-phase integration leading to a universal asymptotic $n-$scaling of the covariances
\begin{equation}
|C^-_{j,k}|  = {\cal O}(n^{-1/2}), \quad {\rm if} \quad j,k\propto n.
\label{eq:rootC}
\end{equation}

\section{Results} 
\begin{figure}[ht!]
\centerline{\includegraphics[angle=-90,width=0.45\textwidth]{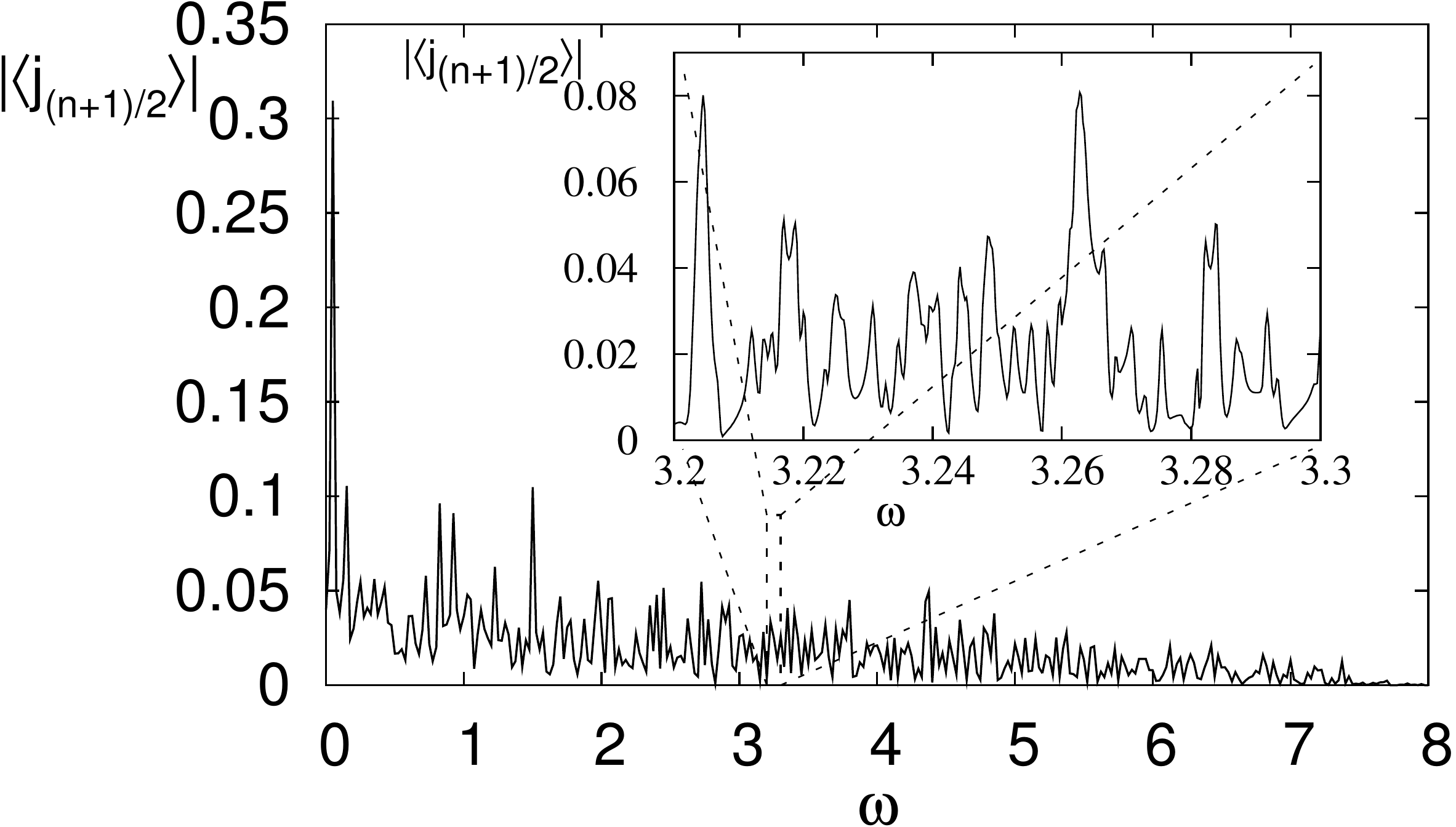}\hskip4mm}
\centerline{\includegraphics[angle=-90,width=0.44\textwidth]{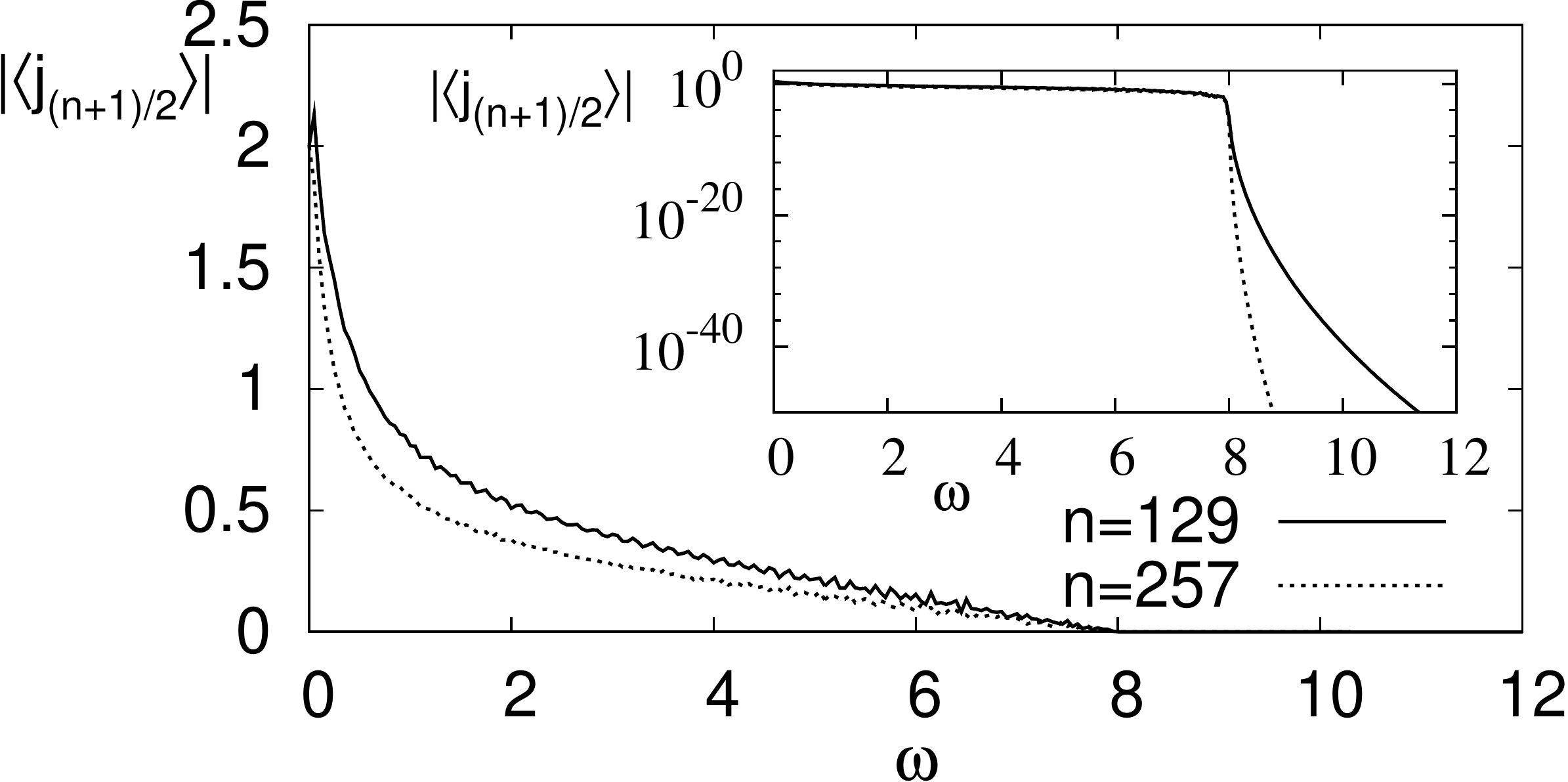}}
\caption{Dependence of the current in the middle of the chain on the driving frequency $\omega$. Top: For small coupling $\varepsilon=0.01, n=257$, many resonances are visible. Bottom: For strong coupling $\varepsilon=1$ we have a smooth dependence. For $\omega> \omega_{\rm c}$ (lower inset) the current decays exponentially with $n$.}
\label{fig:omega}
\end{figure}

\begin{figure}[ht!]
\includegraphics[angle=-90,width=0.245\textwidth]{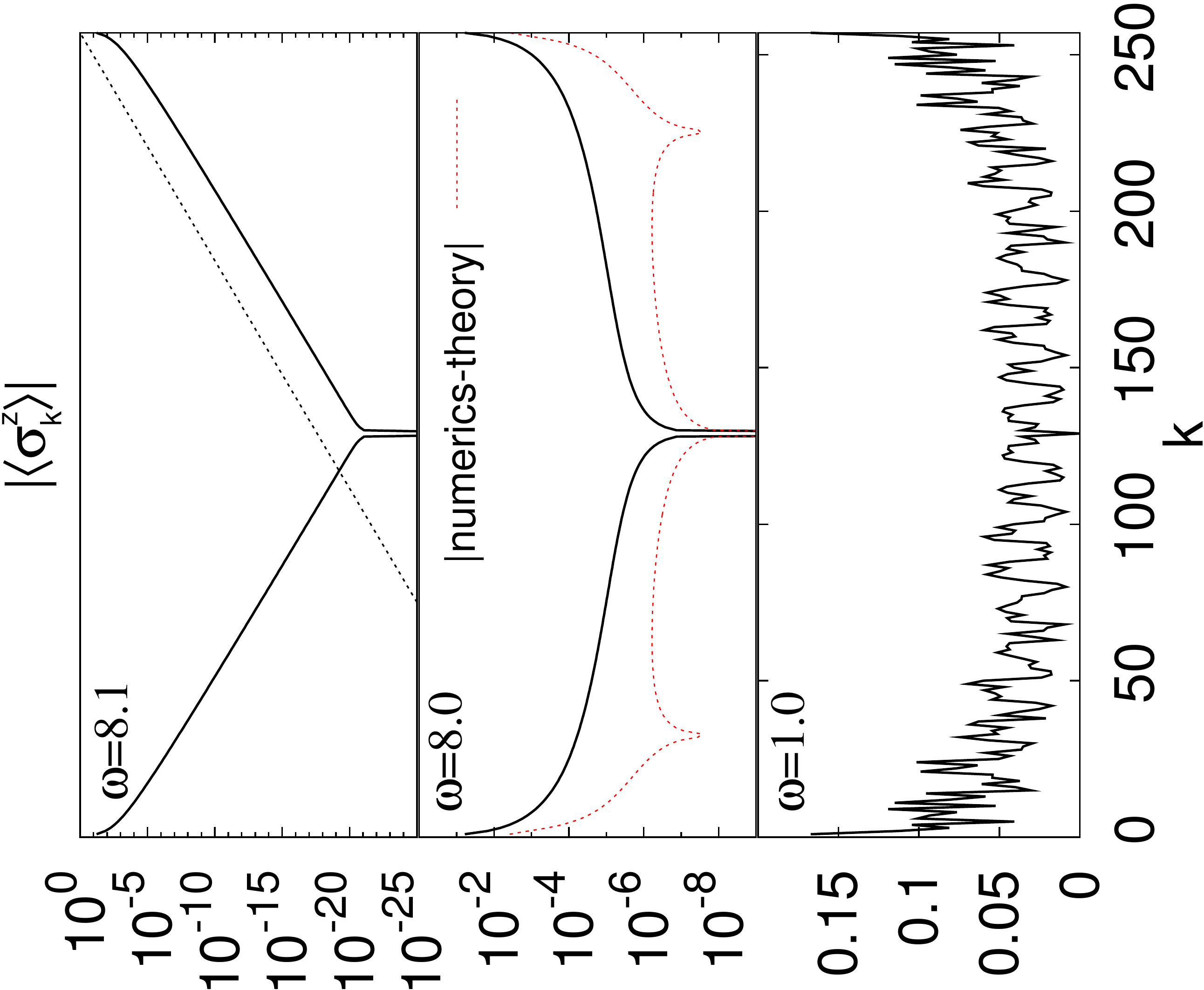}\includegraphics[angle=-90,width=0.245\textwidth]{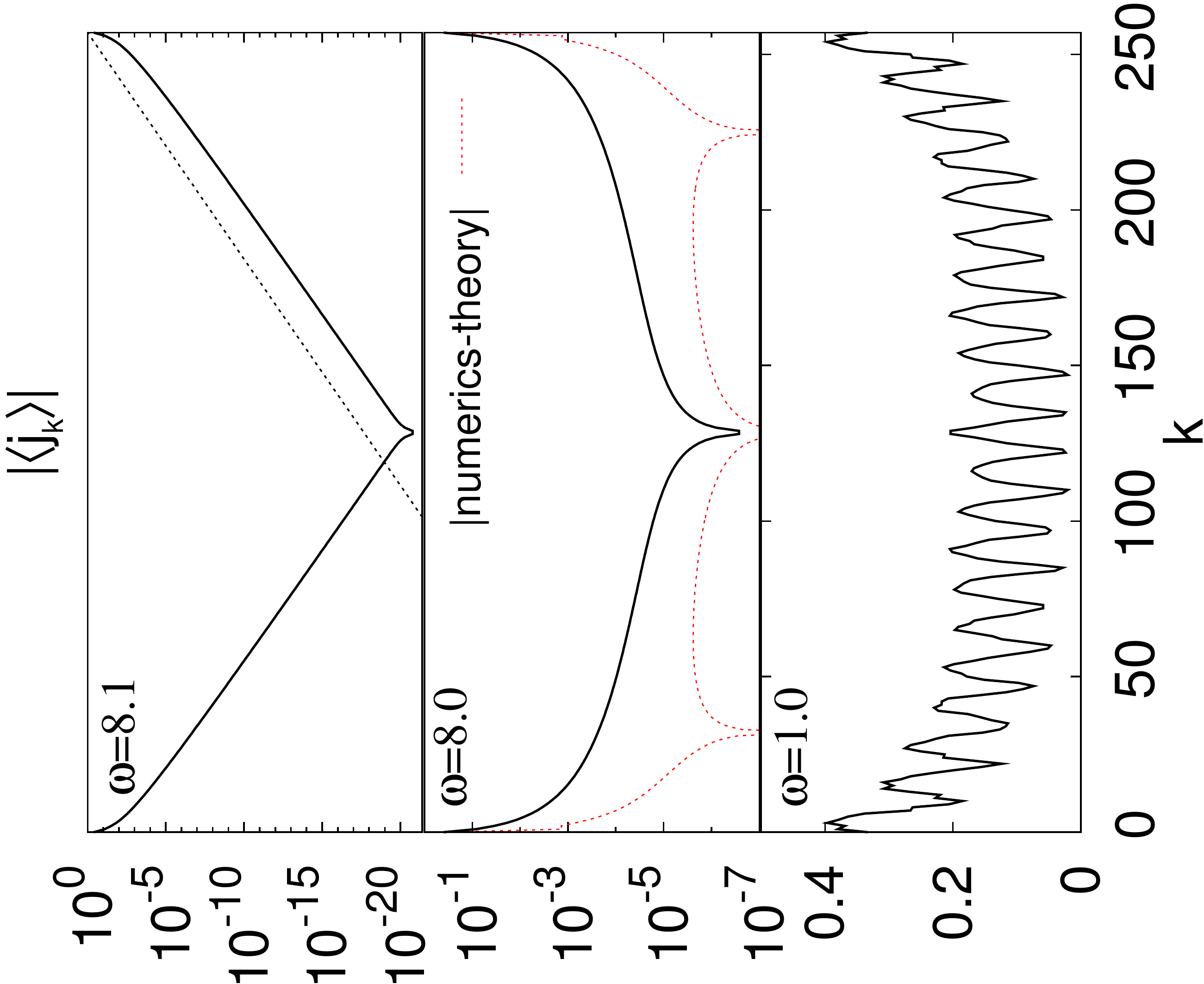}
\caption{(Color online) Magnetization profiles (left column) and spin current profiles (right column) for $n=257$, $\varepsilon=0.1$, and three different $\omega$. At the critical $\omega_{\rm c}=8$, the profiles agree with the theoretical estimate (\ref{eq:Gc}) (relative error is always less than $10\%$). Dashed line for $\omega=8.1$ is $\exp{(-k/\xi)}$ with $\xi = \frac{1}{\sqrt{\omega-\omega_{\rm c}}} \approx 3.16$.}
\label{fig:profiles}
\end{figure}

The main quantity we shall be interested in is the current in the middle of the chain, $j_{\lfloor (n+1)/2\rfloor}$, and its scaling with $n$. The scaling will tell us how much the disturbance from the driving at the chain ends influences the bulk. In Fig.~\ref{fig:omega} we show the current's dependence on the driving frequency $\omega$, which turns out to be the most important parameter in the system, that can qualitatively change the transport. 
For small couplings $\varepsilon$, shown in the top frame of Fig.~\ref{fig:omega} , one can see that overall the current decays with $\omega$, however for $\omega < \omega_{\rm c}$, there are many very narrow resonances. Looking at the theoretical formula (\ref{eq:Cjk}), valid for small $\varepsilon$, it is easy to understand where these resonances come from. They arise from points where the denominator $\lambda_p+\lambda_m$ is very small, which happens when $\omega=\omega_{p-m} \equiv \epsilon_p+\epsilon_m$, with $\epsilon_m=4\cos{a_m}$ being the energies of free fermionic modes. The width of resonances scales as $\Delta \omega \sim \varepsilon/n$, and within the resonance region $|\omega-\omega_{p-m}| < \Delta\omega$ the scaling of covariances is $|C^-_{j,k}|={\cal O}(\varepsilon^0 n^{-1})$ rather than ${\cal O}(\varepsilon^1 n^{-1/2})$ (\ref{eq:rootC}).  For larger couplings (bottom frame in the Fig.~\ref{fig:omega}) the resonances merge into a smooth curve. We can also see that for $\omega > \omega_{\rm c}$ the current decays very rapidly with $\omega$ and $n$, so there the system becomes an insulator for $n \to \infty$. At $\omega=\omega_{\rm  c}$ we have a non-equilibrium phase transition. The value of $\omega_{\rm c}=8$ is also given by the bandwidth of the XX chain or, in physical picture, an insulating regime appears because fermions do not have enough time to reach the neighboring site before the sign of the driving reverses. Next, we show in Fig.~\ref{fig:profiles} the spatial dependence of the magnetization $\ave{\sz_k}$ and the spin current $\ave{j_k}$ along the chain. Note that $\rho(t)$ is time-dependent, even after a long time, so the continuity equation reads $\ii \omega \ave{\sz_k}=\ave{j_{k-1}}-\ave{j_k}$. In addition to the amplitudes, shown in Fig.~\ref{fig:profiles}, there is also a nontrivial dependence of phases on the spatial position $k$ of each of these quantities. We should note that none of the qualitative features depends on the value of $\varepsilon$ (for more see the Appendix~\ref{app:epsilon}). At the transition point $\omega=\omega_{\rm c}$ the correlations can be for large $n$ approximately calculated by summing over only two nearest quadrupoles in (\ref{eq:solC}, \ref{eq:prehod}), resulting in 
\begin{equation}
C_{\rm c}(x,y) \approx \frac{\varepsilon}{n^2}\left( G_{\rm c}(x,y)+(-1)^n G_{\rm c}(x-1,y-1)\right).
\label{eq:Gc}
\end{equation}
The difference between this theoretical dependence and the exact values is shown in Fig.~\ref{fig:profiles} and is small. For $\omega < \omega_{\rm c}$ the main wavelength visible in magnetization and current profiles is determined by the closest resonance.

\begin{figure}[ht!]
\includegraphics[width=0.23\textwidth]{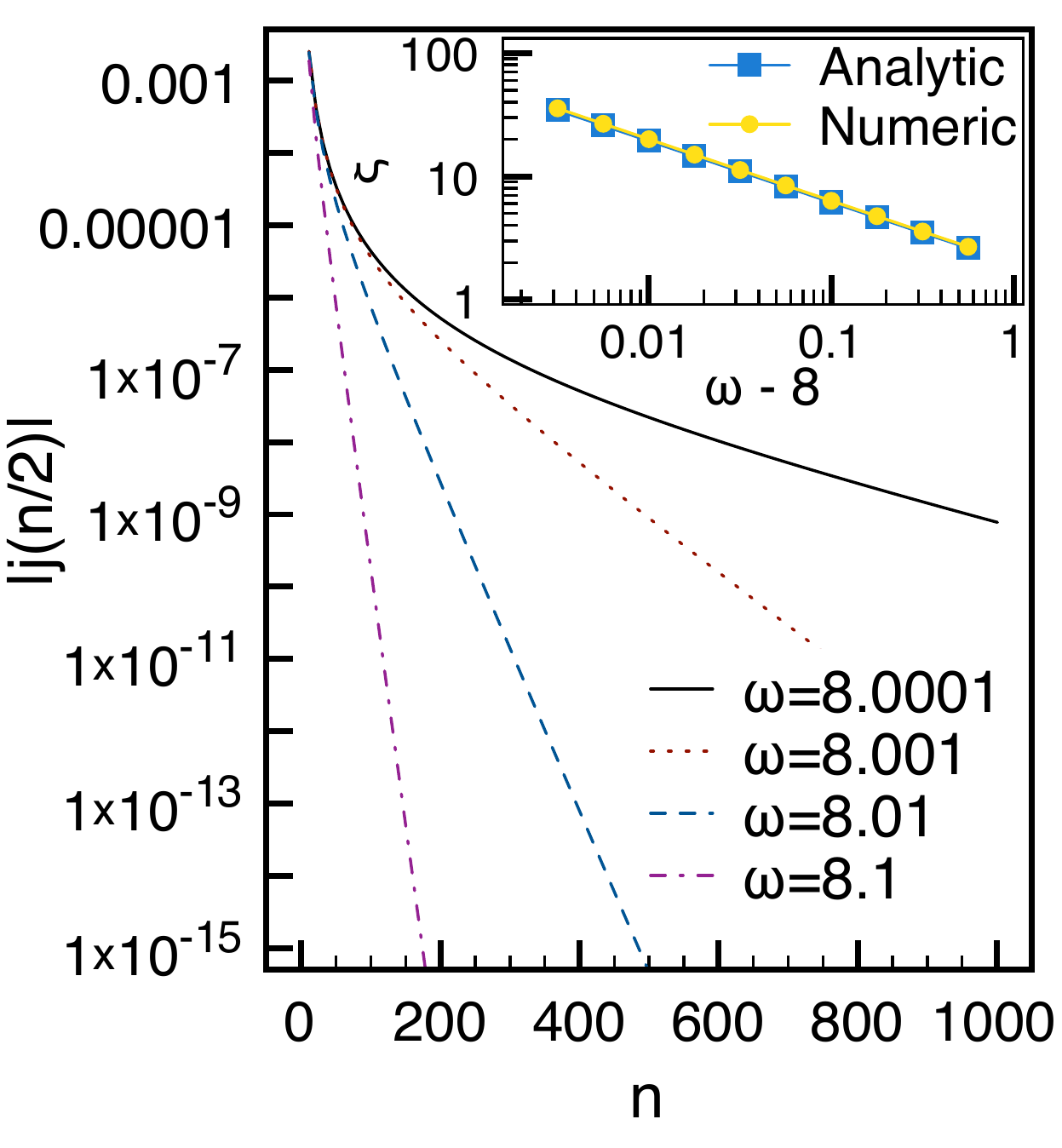} \includegraphics[width=0.23\textwidth]{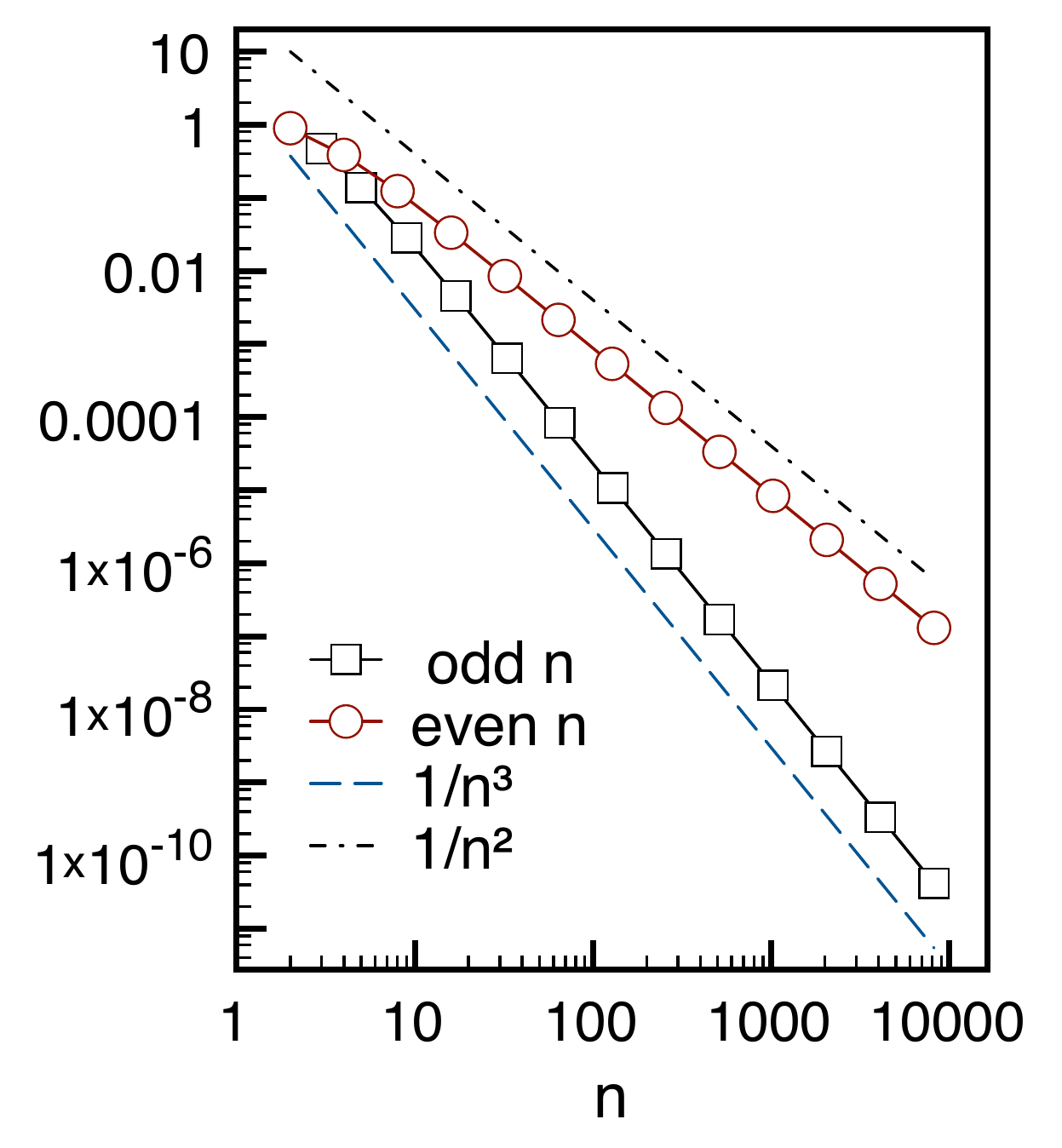} \includegraphics[width=0.46\textwidth]{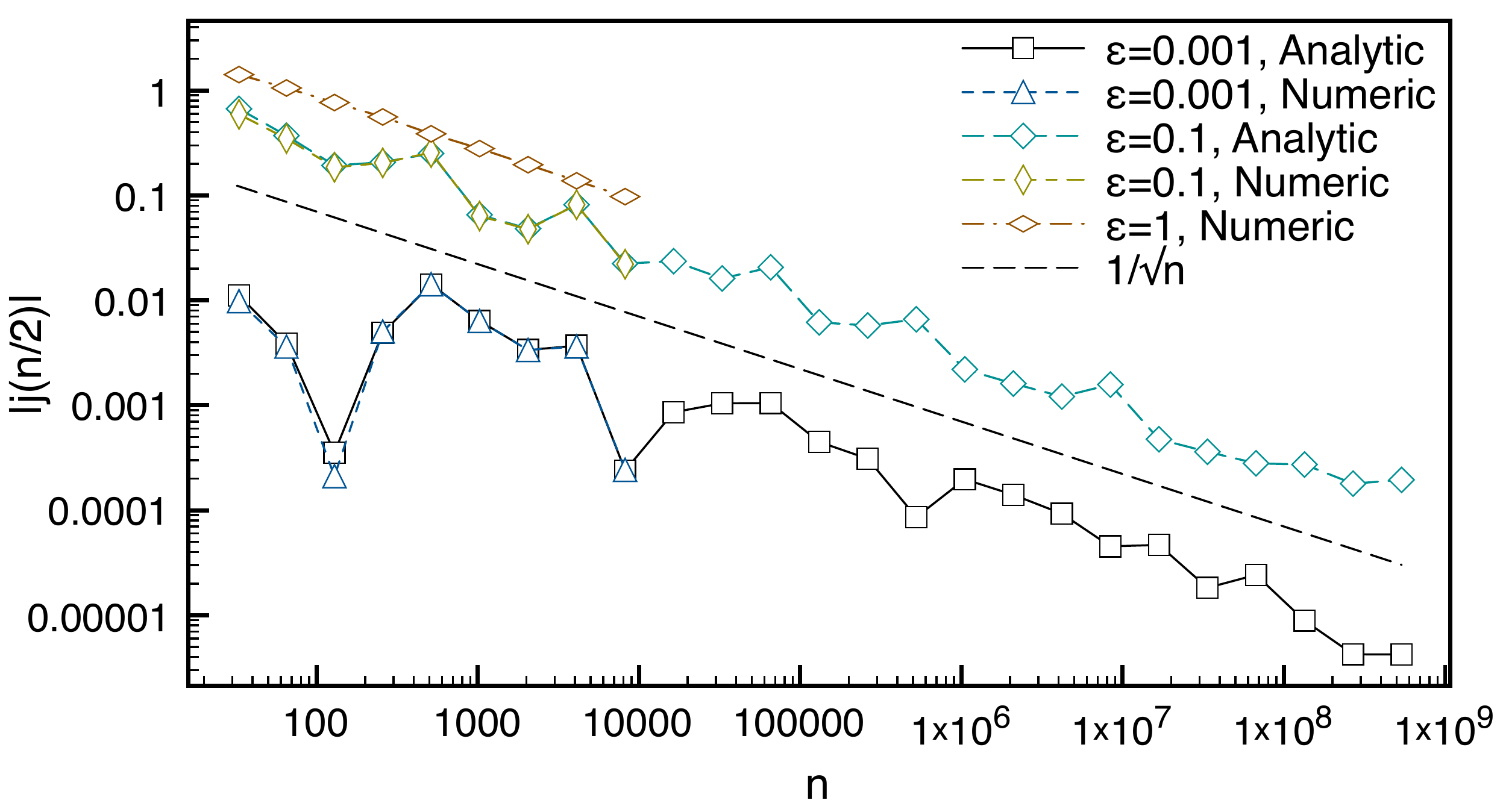} 
\caption{(Color online) Scaling of the current in the middle of the chain on $n$. Top left: $\omega>8$, top right: $\omega=8$, bottom: $\omega=1$. Note that the scaling for the critical $\omega$, being either $\sim 1/n^2$ or $\sim 1/n^3$, depends on the parity of the system size. For $\omega < \omega_{\rm c}$ the dependence on $n$ is for small $\varepsilon$ not monotonic $\sim 1/\sqrt{n}$, however, oscillations get smaller for larger $\varepsilon$; ``Numeric'' points are numerically exact solutions, ``Analytic'' are weak-coupling summation of Eq.~(\ref{eq:Cjk}). For $\omega > \omega_{\rm c}$ the current scales as $\sim \exp{(-\frac{n}{2\xi})}$ with the evanescence length $\xi = \frac{1}{\sqrt{\omega-\omega_{\rm c}}}$ (inset).}
\label{fig:scaling}
\end{figure}

To assess the nature of spin transport at various values of $\omega$ we have studied dependence of the spin current in the middle of the chain on the system size $n$. Above the transition point $\omega> \omega_{\rm c}$ the dependence is exponential, $|j_{(n+1)/2}| \sim \exp{(-\frac{n}{2\xi})}$, as shown in Fig.~\ref{fig:scaling}, with $\xi = \frac{1}{\sqrt{\omega-\omega_{\rm c}}}$, obtained e.g. from Eq.~(\ref{eq:PDE}) (for another approach see the Appendix). At the critical point $\omega=\omega_{\rm c}$ the scaling of the current in the middle of the chain depends on the parity of $n$. For even $n$ it is $\sim 1/n^2$, while it is $\sim 1/n^3$ for odd $n$. This can be theoretically seen from Eq.(\ref{eq:Gc}): for odd $n$ two quadrupoles are subtracted, leading to a subleading-order scaling along the skew-diagonal of $C_{j,k}$. For $\omega=\omega_{\rm c}$ the system is therefore also an insulator in the thermodynamic limit, however, the limit is reached in an algebraic way. In a way, the most interesting regime is for $\omega < \omega_{\rm c}$. If, when one increases $n$ one ``sits'' at a certain resonance (thereby changing $\omega$ with $n$), $\omega_{p-m}=\epsilon_p+\epsilon_m$, and one at the same time also lets $\varepsilon \to 0$, the scaling is $|j_{(n+1)/2}| \sim 1/n$, simply because only the resonance term in Eq.~(\ref{eq:Cjk}) contributes. However, the limit $n \to \infty$ and at the same time $\varepsilon \to 0$, while $\omega$ depends on $n$, is rather artificial. More important is the limit when one fixes $\omega$ and $\varepsilon$, letting only $n \to \infty$. The results, shown in Fig.~\ref{fig:scaling} for extremely large $n \sim 10^9$, indicate that the average decay of the current is $|j_{(n+1)/2}| = 4|C^-_{(n+1)/2,(n+1)/2+1}| \sim 1/\sqrt{n}$, as argued analytically (\ref{eq:rootC}). Such anomalous scaling, although being rather common in classical systems, has been observed only recently in a quantum setting, namely in the isotropic Heisenberg model~\cite{PRL10}. 

\section{Spatial dependence of correlations}

\subsection{Critical $\omega_{\rm c}$}

\begin{figure}[ht!]
\vskip2mm
\includegraphics[width=0.48\textwidth]{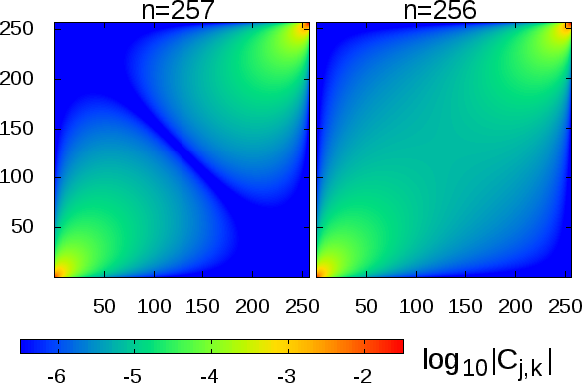}
\caption{(Color online) Exact correlations at the non-equilibrium phase transition point, $\omega=8$. The coupling is $\varepsilon=0.1$. The $C^-_{j,k}$ is approximately given by two quadrupole sources at two opposite corners (high intensity in the figure). For odd $n$ (left figure) two quadrupole sources are subtracted, for even $n$ (right figure) they are added.}
\label{fig:Com8}
\end{figure}

At the critical driving frequency $\omega=\omega_{\rm c}$ the spatial dependence of $C^-_{j,k}$ does not show any oscillations, as is the case for $\omega < \omega_{\rm c}$. In Fig.~\ref{fig:Com8} we show a density plot of $|C^-_{j,k}|$ for $\omega=8$ and $n=257$ as well as $n=256$. 
One can nicely see very high intensity near the two opposite corners, reflecting the fact that the correlations can be approximately described by two quadrupole sources. For odd $n$ these sources are subtracted, resulting in a suppressed correlations along the skew-diagonal, leading there to a $\sim 1/n^3$ scaling. For even $n$ though the quadrupoles are added, resulting in higher correlations scaling as $\sim 1/n^2$ along the skew-diagonal. Also compare these plots with the cross sections along the diagonal (magnetization) and near-diagonal (current) shown in the Fig.~\ref{fig:profiles}.

\subsection{Correlations for $\omega<\omega_{\rm c}$}
\label{sec:Res}

\begin{figure}[ht!] 
\includegraphics[width=0.45\textwidth]{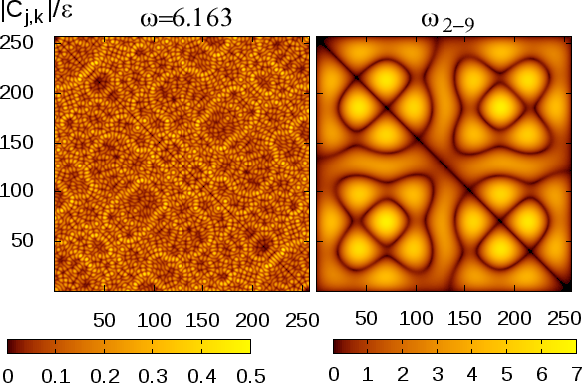} 
\caption{(Color online) Correlations $|C_{j,k}^-|/\varepsilon$ for $n=257$, $\varepsilon=0.001$. On the left is for an out-of-resonance $\omega=6.163$ with an essentially ``random'' structure of $C$; on the right is for $\omega=\omega_{2-9}\approx 7.975$, which is on the resonance, resulting in a simple form $C \sim |\sin{(2\pi x)}\sin{(9\pi y)}+\sin{(9\pi x)}\sin{(2\pi y)}|$.}
\label{fig:C}
\end{figure}

In Fig.~\ref{fig:C} we show the expectations of all nonzero 2-point observables, i.e., $C^-_{j,k}$, at smaller frequency than $\omega_{\rm c}$.  
We can see that the structure of correlations can be quite different at different frequencies.
\begin{figure}[ht!]
\includegraphics[angle=-90,width=0.47\textwidth]{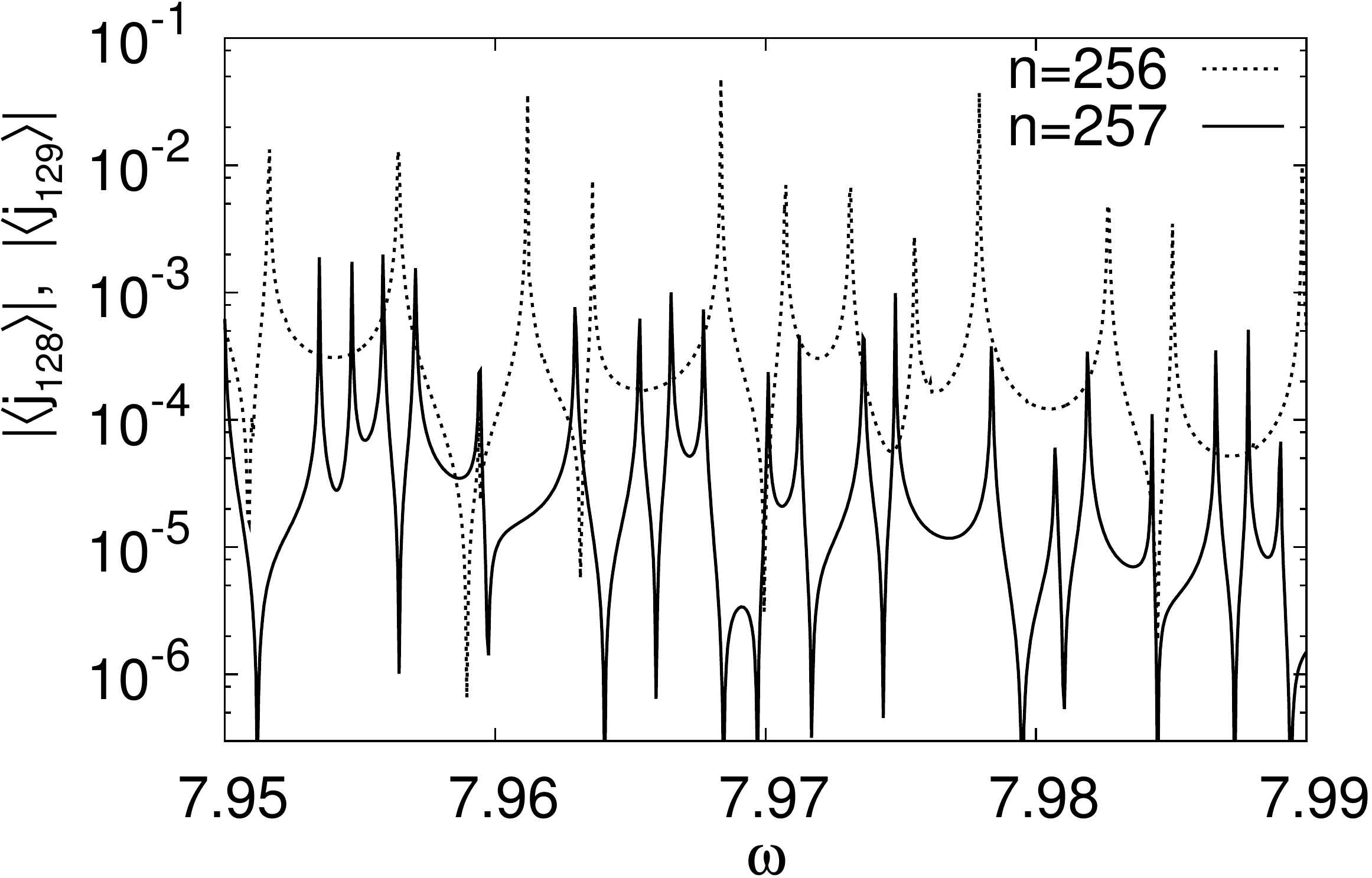}
\caption{Resonances are for odd and even $n$ at different positions. Current at the middle of the chain is for $n=256$ by two orders of magnitude larger than for $n=257$ because we are close to $\omega_{\rm c}=8$, where one scales as $\sim 1/n^2$, while the other is $\sim 1/n^3$. Both data are for $\varepsilon=0.01$.}
\label{fig:lihisodi}
\end{figure} 
Below the transition point and for small $\varepsilon$ the correlations are dominated by resonances, for which the condition $\omega=\omega_{p-m}=\epsilon_p+\epsilon_m$ ($\epsilon_p=4\cos{\frac{p\,\pi}{n+1}}$) is fulfilled. As explained, the position of resonances is different for even and odd sizes $n$ because in one case only even $p+m$ are allowed while in the other only odd $p+m$ occur. This is illustrated in Fig.~\ref{fig:lihisodi}.

\begin{figure*}[ht!]
\centerline{\includegraphics[width=\textwidth]{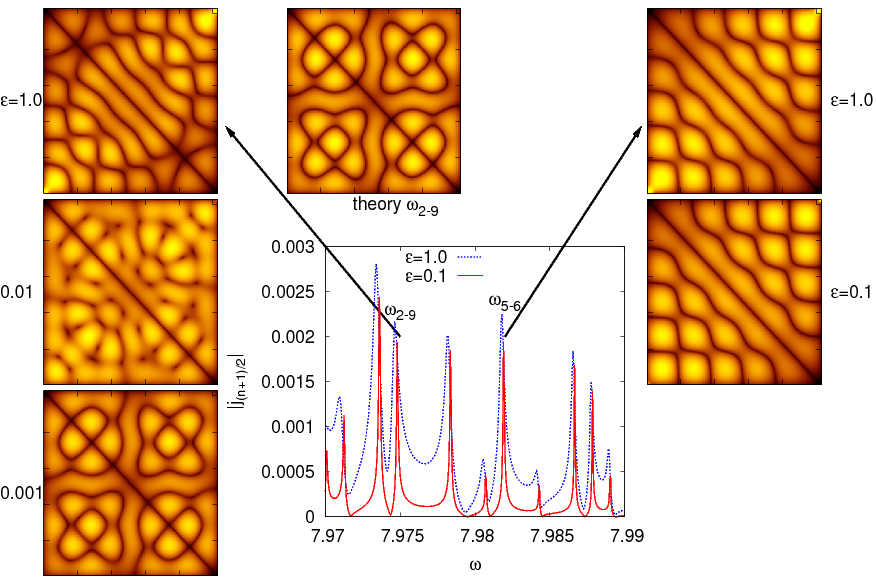}}
\caption{(Color online) Transition to the theoretical correlation function (\ref{eq:Cth}) at high resonant frequencies. On the left a case of the $2-9$ resonance is shown, while the right pictures show the $5-6$ resonance. Top middle picture shows the theoretical resonance shape, Eq.~(\ref{eq:Cth}), for $p=2$, $m=9$. See text for more details. Color scale is adjusted between the plots for better detail, all is for $n=257$.}
\label{fig:Call}
\end{figure*}

If the driving frequency $\omega$ is equal to some resonant frequency $\omega_{p-m}$, where $p+m=n \,({\rm mod}\, 2)$, and if $\varepsilon$ is sufficiently small, the spatial pattern of correlations is given by the theoretical formula
\begin{equation}
|C_{j,k}^-| \propto |\sin{(p\pi x)}\sin{(m\pi y)}+\sin{(m\pi x)}\sin{(p\pi y)}|.
\label{eq:Cth}
\end{equation}
This form comes due to a combination of two eigenfunctions. How small must the coupling be for this to happen is a nontrivial question. It depends on how well is the resonance in question resolved. For instance, the density of resonances is in general higher at smaller $\omega$. This means that in order to resolve a resonance one will typically need a smaller $\varepsilon$ at smaller frequencies than for instance just below the critical frequency. We illustrate this phenomenon in Fig.~\ref{fig:Call}. Two resonances are shown: the resonance $5-6$ at $\omega_{5-6}\approx 7.98192$ is well separated from the others, while the $2-9$ resonance at $\omega_{2-9}\approx 7.97482$ is very close to the $\omega_{6-7}\approx 7.97481$. In fact, in the plot of $|j_{(n+1)/2}|$ on $\omega$ these two resonances can not be resolved on the scale of the plot. Because of that, at large $\varepsilon=1$ the spatial dependence of correlations is a kind of merger of resonances $\omega_{2-9}$ and $\omega_{6-7}$. The theoretical small-$\varepsilon$ shape of the resonance is resolved only around $\varepsilon=0.001$. Spatial pattern of the resonance $\omega_{5-6}$ is on the other hand well resolved already at large $\varepsilon=1$.

\begin{figure}[ht!]
\includegraphics[width=0.49\textwidth]{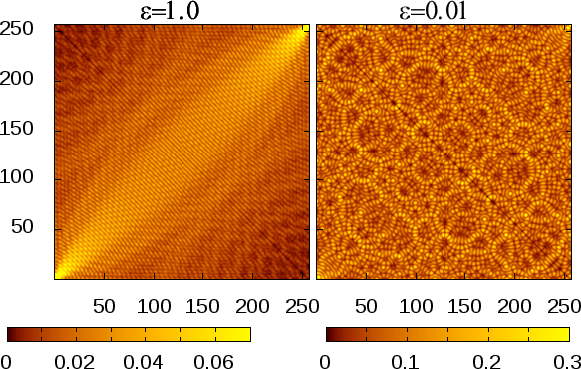}
\vskip2mm
\includegraphics[width=0.49\textwidth]{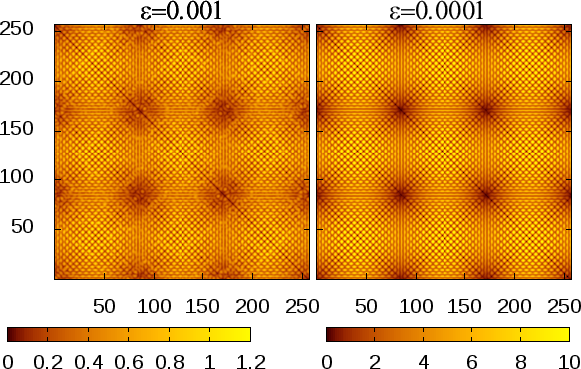}
\caption{(Color online) Resonance $\omega_{3-82}$ at various $\varepsilon$. This resonance is the closest to $\omega=6.163$ shown in Fig.~\ref{fig:C}. We plot $|C_{j,k}|/\varepsilon$.}
\label{fig:Cres3-82}
\end{figure}

For small $\omega$ the transition to theoretical $|C_{j,k}^-|$ typically happens at smaller couplings. This is shown in Fig.~\ref{fig:Cres3-82} where we show the $3-82$ resonance. This resonance is the closest to the frequency $\omega=6.163$, whose correlations have been shown in Fig.~\ref{fig:C}. We can see that one needs $\varepsilon = 0.0001$ in order to reach Eq.(\ref{eq:Cth}). The shape of correlations at general small frequencies and large couplings typically looks rather ``random'', similar to plots at $\varepsilon=1.0$ or $\varepsilon=0.01$ in Fig.~\ref{fig:Cres3-82} .

\section{Other reservoirs}
The physical phenomenon described for our exactly solvable model, i.e. a phase transition with $\omega$, is robust to small changes of the model. For instance, in this Section we show that similar transition is obtained also for a system coupled to the so-called two-spin baths, in each of which one has $16$ Lindblad operators acting on two boundary sites (simulating finite-''temperature''). Physical picture, explaining why the phase transition occurs, is clear. If driving is faster than the fastest time scale in the systems -- the nearest neighbor coupling (hopping strength) -- then the system is an insulator. We therefore conjecture that the phenomenon is robust to, for instance, changes of the reservoir. The only necessary ingredient is that one has reservoirs of magnetization, or, in fermionic picture, of particles. The advantage of reservoirs studied in the present work, i.e., with Lindblad operators $L \sim \sigma^\pm$, is that we are able to analytically solve for the nonequilibrium steady state. For other types of reservoir Lindblad operators the system is not quadratic in fermionic variables anymore and exact solution is in general not possible. Nevertheless, to test our conjecture we performed numerical simulations with time-dependent density matrix renormalization group (tDMRG) method. We used the so-called two-spin reservoirs, in which there are $16$ Lindblad operators acting on two spins at each chain end. Details of our implementation can be found in Ref.~\cite{tdmrg}.
\begin{figure}[ht!]
\includegraphics[angle=-90,width=0.49\textwidth]{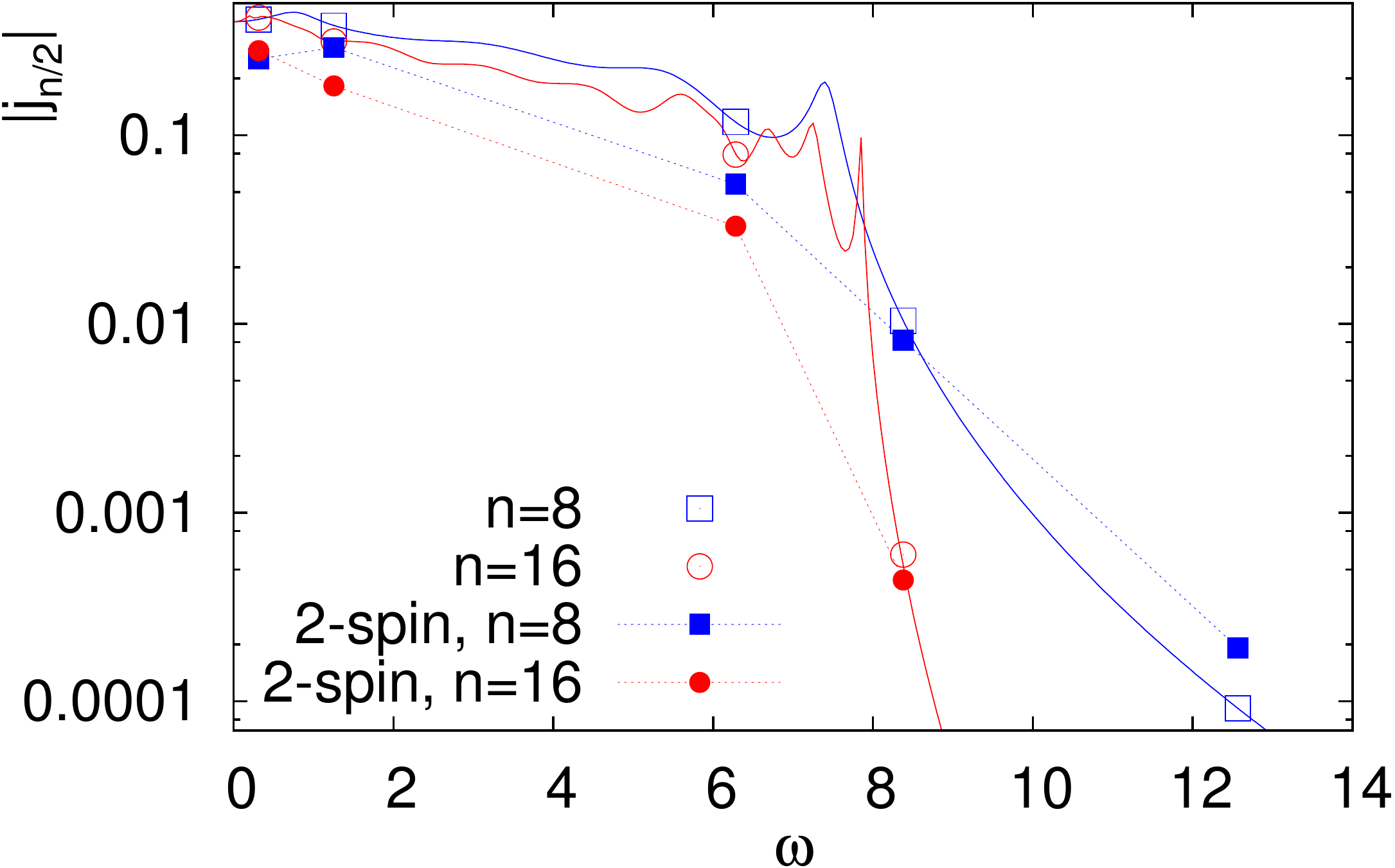}
\caption{(Color online) Spin current in the middle of the chain for different frequencies. Symbols, squares for $n=8$ and circles for $n=16$, are results of tDMRG simulations, two full lines are analytic results for $L \sim \sigma^\pm$. Open symbols are tDMRG simulation of the system studied in the present work ($L \sim \sigma^\pm$) while full symbols are for a two-spin bath, simulating finite-temperature reservoirs. $\varepsilon=1$, $\mu_0=0.2$.}
\label{fig:finiteT}
\end{figure}
Because experimental realization, for instance with ions, is easier for short chains we choose a short chain of length $n=8$ and $n=16$, in order to demonstrate that the phase transition can be seen already in small systems. Results of numerical simulations are shown in Fig.~\ref{fig:finiteT}. First, we cross-checked our analytical results for Lindblad operators $L \sim \sigma^\pm$ presented in this work with the results of tDMRG simulations (open symbols). By comparing two points at $\omega \approx 6$ and $\omega \approx 8.4$ one can see a large drop in the current around $\omega_{\rm c}=8$. Using a two-spin bath one can achieve a nonequilibrium steady state with nonzero average energy density. Assuming that the state is locally close to a canonical one, one can ascribe temperature to the nonequilibriums steady state by comparing the average energy density with the canonical one~\cite{thermal}. In our simulations (full symbols in Fig.~\ref{fig:finiteT}) the energy density in the steady state is $\ave{\sigma_j^{\rm x} \sigma_{j+1}^{\rm x} +\sigma_j^{\rm y} \sigma_{j+1}^{\rm y}} \approx -0.26$, which corresponds to the canonical expectation value at temperature $T \approx 7.6$. The nonequilibrium steady state of the analytical solution has on the other hand zero average energy density and can be described as a state at an infinite temperature. Observing data points for two-spin bath and $n=16$ (full circles) at $\omega \approx 6$ and at $\omega \approx 8.4$ we can see that even for a non-solvable finite-temperature NESS there is still a two orders of magnitude drop in the current as one increases the driving frequency beyond $\omega_{\rm c}$. Phase transition is therefore quite robust and is not only a property of our solvable model.

\section{Experimental implementation}

Relatively novel way of implementing various quantum models is via a rapidly developing field of simulating physical systems with cold atoms or ions. All ingredients necessary to implement our model, like the exchange interaction between nearest neighbors, have already been achieved~\cite{cold}. Simulating Lindblad equation, for instance in order to dissipatively prepare a given pure state~\cite{dissipative-state,open-exper}, is by now also quite established. We shall here sketch the implementation with ions, although other realizations, for instance with atoms in an optical lattice~\cite{digital-atoms}, go along similar lines. The method would actually be very similar to the one used in Ref.~\cite{open-exper,digital-ions} where a master equation with the Lindblad operator $L'=\frac{1}{2}\sigma_1^{\rm z}(\mathbbm{1}-\sigma_1^{\rm x}\sigma_2^{\rm x}\sigma_3^{\rm x}\sigma_4^{\rm x})$ has been implemented on a system of 4 ions. In our model we instead need $L=\sigma_1^+$. Difference from Refs.~\cite{open-exper,digital-ions} would be that to implement $L$ one has to apply an entangling M\o lmer-S\o rensen gate~\cite{MS} just on 2 ions (ancilla and one system's site) instead of on 5 ions -- for description see e.g. the appendix B. in Ref.~\cite{digital-ions}. This can be seen by writing $L=\sigma_1^+=\frac{1}{2}\sigma_1^{\rm x}(\mathbbm{1}-\sigma^{\rm z}_1)$. Up to a trivial rotation around $y$-axis this is similar to $L'$. Phase transition could be detected by simply measuring the state (magnetization) of ions. Another feasible way to implement our model is in a mesoscopic setting with electrons~\cite{Kohler}. One would need a quantum wire (coupled quantum dots, molecular wire, etc.) coupled to electron reservoirs. By varying electro-chemical potential in the reservoirs, for instance by an external gate potential or a laser pulse, one can achieve a time-dependent occupation of reservoirs, which would correspond to our magnetization reservoirs in the spin language. In fact, a somewhat related model of tight-binding electrons in an a.c. electric field~\cite{fieldE} has been studied extensively by approximate theoretical and numerical methods, see e.g.~\cite{arrays} and references therein. 

\section{Conclusion} We have analytically solved a system of 1D spinless fermions under harmonic a.c. driving at the lattice ends. With the driving frequency there is a transition from a system with a ballistic transport for $\omega=0$, to the one with anomalous transport for $\omega<8$, which at the critical frequency $\omega=8$ changes to an insulator. 

We acknowledge fruitful discussions with M. Horvat and support by the grants P1-0044 and J1-2208 of Slovenian Research Agency (ARRS), and thank T.~H.~Seligman for reading the manuscript.

\appendix
\section{Dependence on the coupling strength}
\label{app:epsilon}

\begin{figure}[ht!]
\includegraphics[angle=-90,width=0.45\textwidth]{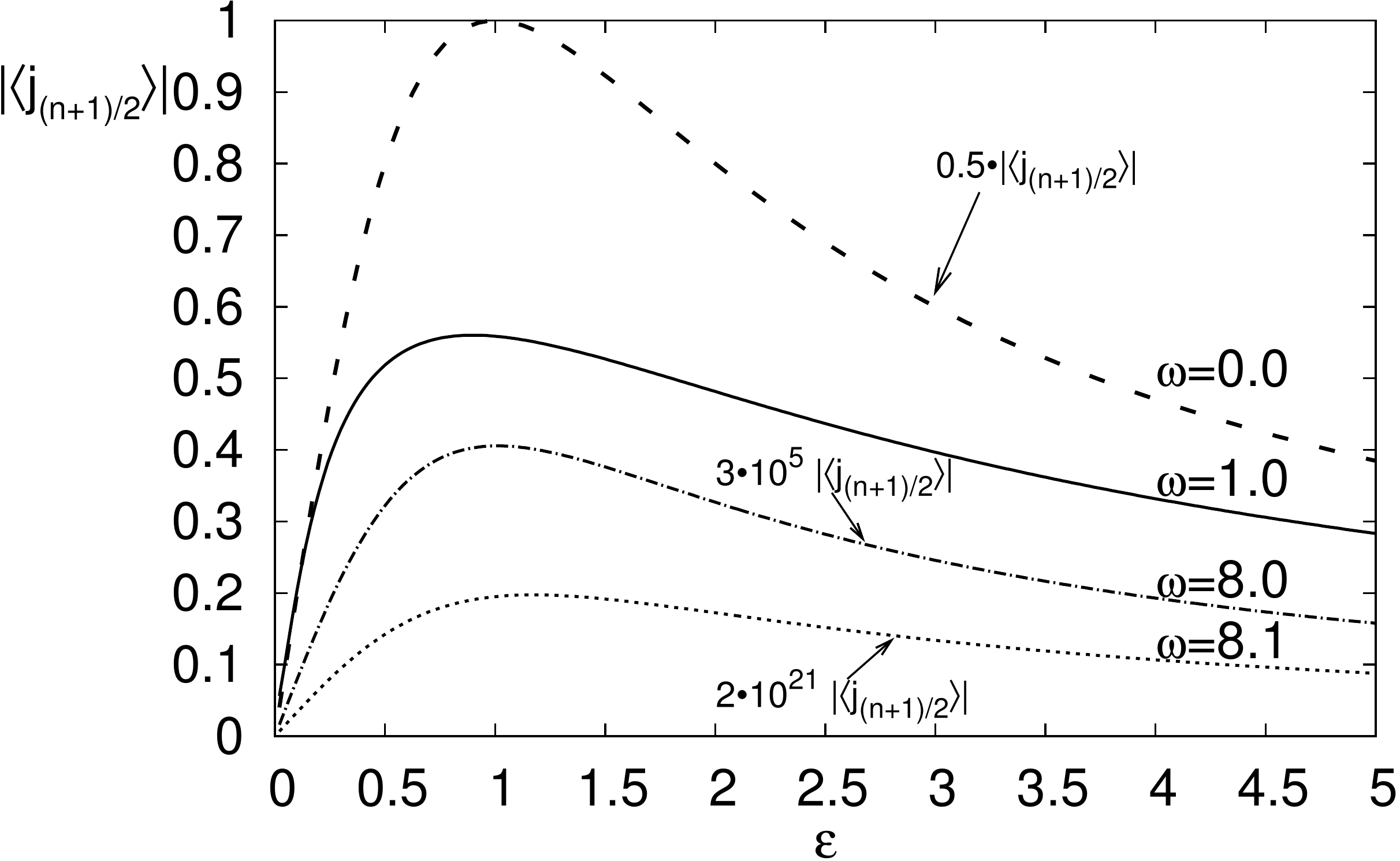} 
\caption{Dependence of the current in the middle of the system on the coupling strength $\varepsilon$. For all $\omega$ the overall functional dependence is similar to the one at $\omega=0$ (dashed curve), only the size of the current changes (note different prefactors for $\omega \neq 1.0$); all is for $n=257$.}
\label{fig:Gamma}
\end{figure}

The coupling strength $\varepsilon$ between baths and the system is a rather trivial parameter that does not affect the main qualitative features of the transport. It of course influences the size of correlations $C_{j,k}$. In Fig.~\ref{fig:Gamma} we show the dependence of the current in the middle of the chain, $j_{(n+1)/2}$ on $\varepsilon$. The functional form is always very similar to the one at $\omega=0$, which is known exactly~\cite{Karevski:09,JPA:10}, and is $\ave{j_{k}}=\frac{4}{\varepsilon+\frac{1}{\varepsilon}}$. 
More important is the influence of the value of $\varepsilon$ on the width of resonances and whether the latter are isolated or not, which is in turn reflected in the spatial pattern of $C_{j,k}$.

\section{Exact solution of the weak coupling limit}
\label{sec:pert}

The stationary correlation matrix ${\bf C}^-$ is obtained from the continuous Lyapunov equation
\begin{eqnarray}
\label{eq:lyap2}
\{{\bf A},{\bf C}^{-}\}=-4\varepsilon {\bf S},\quad{\bf A}=2({\bf J}+{\rm i}\varepsilon {\bf R})-\frac{\omega}{2}\mathbbm{1},
\end{eqnarray}
with ${\bf J}$ being an $n\times n$ matrix with the only nonzero elements $J_{k,k+1}=J_{k+1,k}=1$ while the $n\times n$ matrices ${\bf S,~R}$ have all elements equal to zero except $R_{1,1}=R_{n,n}=S_{1,1}=(-1)^{n}S_{n,n}=1$. The matrix ${\bf A}$ consists of a term ${\bf J}$, which corresponds to kinetic energy of the system, a coupling term ${\bf R}$, that corresponds to the dissipation to the environment, and one additional  term, that comes from the time dependent, oscillatory part of the correlation matrix. The source term ${\bf S}$ is determined by the driving, i.e. the difference in the chemical potentials between the left and right reservoirs. The equation (\ref{eq:lyap2}) is straightforwardly solved by the following ansatz
\begin{eqnarray}
\label{eq:ansatz}
\mathbf{C}^-=\sum_{j,k}\Lambda_{j,k}\underline{\Psi}_j^\mathrm{right}\otimes\underline{\Psi}_k^\mathrm{right},
\end{eqnarray}
where $\underline{\Psi}_j^\mathrm{right}$ denotes the $j$-th right eigenvector of $\mathbf{A}$ with the corresponding eigenvalue $\beta_j$, $\mathbf{A}\underline{\Psi}_j^\mathrm{right}=\beta_j\underline{\Psi}_j^\mathrm{right}$. Then, by plugging the ansatz into the equation (\ref{eq:lyap2}) we get the expression for the coefficients $\Lambda_{j,k}$
\begin{equation}
\label{eq:coef}
\Lambda_{j,k}=-\frac{4\varepsilon}{\beta_j+\beta_k}\underline{\Psi}_j^{\mathrm{left}*}\cdot\mathbf{S}\underline{\Psi}_k^{\mathrm{left}*}.
\end{equation}
We find the eigenvectors and the eigenvalues perturbatively in $\varepsilon$ and calculate only the leading contribution to the correlation matrix ${\bf C}^-$. The leading-order eigenvectors and the corresponding eigenvalues are
\begin{eqnarray}
\label{eq:0order}
\psi^{(0)}_{j,k}=\sqrt{\frac{2}{n+1}}\sin a_{jk},\quad
\beta^{(0)}_{j}=\frac{\omega}{2}-4 \cos a_j,
\end{eqnarray}
where $a_k \equiv\frac{\pi k}{n+1}$. For $\omega<\omega_{\rm c}\equiv 8$ the sum of the eigenvalues in the denominator of the ansatz (\ref{eq:ansatz}) can vanish, hence we need to take into account the first order correction to the eigenvalues as well
\begin{eqnarray}
\label{eq:1order}
\beta_{j}^{(1)}=-\frac{8{\rm i}\varepsilon}{ (n+1)}\sin ^2 a_j.
\end{eqnarray}
Combining the equations (\ref{eq:ansatz}), (\ref{eq:coef}), (\ref{eq:0order}), and (\ref{eq:1order}) we immediately get the correlation matrix in the leading order in $\varepsilon$,
\begin{equation}
C_{j,k}^-=-32\varepsilon\!\!\!\!\!\sum_{\substack{p,m=1\\p+m=n ({\rm mod\,} 2)}}^n\!\!\frac{\sin{a_p}\sin{a_m}\sin{a_{jp}}\sin{a_{k m}}}{(n+1)^2(\lambda_p+\lambda_m)},
\label{eq:Cjk_sup}
\end{equation}
where $\lambda_m=\beta^{(0)}_m+\beta^{(1)}_m$.

\section{Exact solution of the weak coupling limit for $\omega\geq\omega_{\rm c}$}
\label{app:hyper}

In the case $\omega>\omega_{\rm c}$ the ansatz (\ref{eq:ansatz}) has no singularities and becomes analytic also in the zeroth order in $\varepsilon$, i.e. by taking $\lambda_m\approx \beta_m^0$, and for all system sizes $n$. Therefore, we can find the solution of the equation (\ref{eq:lyap2}) using a different approach. First we simplify the continuous Lyapunov equation (\ref{eq:lyap2})
\begin{equation}
\label{eq:lyap3}
2\{{\bf J},{\bf C}^{-}\}-\omega{\bf C}^{-}=-4\varepsilon {\bf S}.
\end{equation}
The exact solution of the equation (\ref{eq:lyap3}) is sought in the form of a perturbative ansatz
\begin{equation}
\label{eq:corr_exp}
{\bf C}^-=\sum_{j=0}^\infty\frac{1}{\omega^j}{\bf C}_j.
\end{equation}
The straightforward recursive solution is
\begin{eqnarray}
\label{eq:rec}
{\bf C}_j=\{2{\bf J},{\bf C}_{j-1}\}, \quad {\bf C}_0=\frac{4\varepsilon\bf S}{\omega}.
\end{eqnarray}
From the recursion (\ref{eq:rec}) it is possible to calculate each term in the solution (\ref{eq:corr_exp}) explicitly. After a tedious calculation we then rewrite the correlations as follows
\begin{equation}
\label{eq:sol_exp}
C_{j,k}^-= 4\varepsilon \sum_{l,m=-\infty}^{\infty}\sum_{\nu=0}^1 \frac{(-1)^{\nu n}}{n^2}\, G_{j-2l-\nu,k-2m-\nu},
\end{equation}
where
\begin{eqnarray}
\label{eq:green2}
G_{j,k}=j k \omega ^{-j-k} \Gamma (j+k-1) \Gamma (j+k+1) \,
   \times \\~~~~~~ \nonumber _4\tilde{F}_3\left[ \begin{array}{c} \frac{j+k-1}{2} ,\frac{j+k}{2} ,\frac{j+k+1}{2},\frac{j+k+2}{2}\\ j+1,k+1,j+k+1\end{array} ;\frac{16}{\omega ^2}\right],
\end{eqnarray}
by means of the standard Gamma function $\Gamma(x)$ and generalized hypergeometric function $_4\tilde{F}_3$. Note that for large system sizes $n\gg 1$ and for elements of the covariance matrix lying near the diagonal $|j-k|\ll n$ we can approximate
\begin{equation}
\label{eq:approx1}
C_{j,k}^-\approx\frac{\varepsilon}{n^2} \left(G_{j,k}+(-1)^{n}\, G_{j-1,k-1}\right).
\end{equation}
From the approximate solution ({\ref{eq:approx1}}) it is possible to extract the scaling of the elements near the diagonal in the limit $j\gg1$. We find that $C_{j,j+1}^-\propto e^{-j/\xi}$, where $\xi=\frac{1}{\sqrt{\omega-\omega_{\rm c}}}$.


\begin{thebibliography}{}
\bibitem{footnote3} For $\omega>8$ the real part in $\lambda_p+\lambda_m$ is always positive and much larger than the imaginary part which is $\propto \frac{\varepsilon}{n+1}$.
\end{thebibliography}
\end{document}